\documentclass[aps,preprint]{revtex4}%
\usepackage{amsfonts}
\usepackage{amsmath}
\usepackage{amssymb}
\usepackage{graphicx}%
\begin{document}
\preprint{ }
\title{Spectral convexity for attractive $SU(2N)$ fermions}
\author{Dean Lee}
\affiliation{Department of Physics, North Carolina State University, Raleigh, NC 27695-8202}

\begin{abstract}
We prove a general theorem on spectral convexity with respect to particle
number for $2N$ degenerate components of fermions. The number of spatial
dimensions is arbitrary, and the system may be uniform or constrained by an
external potential. \ We assume only that the interactions are governed by an
$SU(2N)$-invariant two-body potential whose Fourier transform is negative
definite. \ The convexity result implies that the ground state is in a
$2N$-particle clustering phase. \ We discuss implications for light nuclei as
well as asymmetric nuclear matter in neutron stars.

\end{abstract}
\keywords{Inequalities, nuclear matter, effective field theory, lattice, Wigner symmetry}
\pacs{13.75.Cs, 21.30.-x, 21.65.+f, }
\maketitle

Interacting fermions with more than two components exhibit a variety of low
temperature phenomena. \ Of particular interest are phenomena which appear in
different quantum systems and therefore could be characterized as universal.
\ One example in three dimensions is the Efimov effect, which predicts a
geometric sequence of trimer bound states for interactions in the limit of
zero range and infinite scattering length
\cite{Efimov:1971a,Efimov:1993a,Bedaque:1998kg,Bedaque:1998km,Bedaque:1999ve,Braaten:2004a}%
. \ The Efimov effect is forbidden for two-component fermions due to the Pauli
exclusion principle but can occur for more than two components. \ Efimov
trimers have recently been observed in ultracold cesium as indicated by a
large three-body recombination loss near a Feshbach resonance
\cite{Kraemer:2006Nat}. Once the binding energy of the trimer system is fixed
for interactions at zero range and large scattering length, the binding energy
of the four-body system is also determined. \ This is in direct analogy with
the Tjon line relating the nuclear binding energies of $^{3}$H and $^{4}$He
\cite{Platter:2004pra,Platter:2004zs,Hammer:2006ct}. \ In two dimensions a
different geometric sequence has been predicted for zero-range attractive
interactions. \ In this case the geometric sequence describes the binding
energy of $N$-body clusters as a function of $N$ in the large $N$ limit
\cite{Hammer:2004x, Platter:2004x, Blume:2004, Lee:2005xy}.

Several recent studies have investigated pairing and the superfluid properties
of three-component fermions \cite{Bedaque:2006three, He:2006PRA, Zhai:2006,
Paananen:2006PRA}. \ Systems involving four-component fermions are of direct
relevance to the low-energy effective theory of protons and neutrons. \ Due to
antisymmetry there are only two S-wave nucleon scattering lengths,
corresponding with the spin-singlet and spin-triplet channels. \ Some general
properties of this low-energy effective theory have been studied such
as\ pairing, the fermion sign problem, and spectral inequalities
\cite{Lee:2004ze,Lee:2004hc,Chen:2004rq,Wu:2005PRB}. \ Wu and collaborators
\cite{Wu:2003PRL, Wu:2005PRL, Wu:2006MPLB} have pointed out that the effective
theory has an accidental $SO(5)$ or $Sp(4)$ symmetry, and several different
phases such as quintet Cooper pairing or four-fermion quartetting could be
experimentally realized for different scattering lengths with ultracold atoms
in optical traps or lattices \cite{Lecheminant:2005PRL,Capponi:2006}. \ When
the scattering lengths are equal the symmetry is expanded to $SU(4)$. \ This
symmetry was first studied by Wigner \cite{Wigner:1939a} and arises naturally
in the limit of large number of colors for quantum chromodynamics
\cite{Kaplan:1995yg,Kaplan:1996rk}. \ The fact that both the spin-singlet and
spin-triplet nucleon scattering lengths are unusually large means that the
physics of low-energy nucleons is close to the Wigner limit
\cite{Mehen:1999qs,Epelbaum:2001fm}.

\begin{figure}
[ptb]
\begin{center}
\includegraphics[
height=2.8746in,
width=2.1715in
]%
{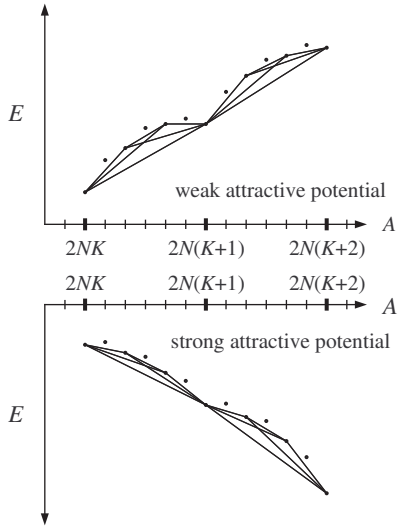}%
\caption{Illustration of the convexity constraints for the ground state energy
$E$ as a function of particle number $A$. \ The line segments show the
convexity lower bounds.}%
\label{convex}%
\end{center}
\end{figure}

In the following we prove a general theorem on spectral convexity with respect
to particle number for $2N$ degenerate components of fermions. \ The theorem
holds for any number of spatial dimensions, and the system may be either
uniform or constrained by an external potential. \ We assume only that the
interactions are governed by an $SU(2N)$-invariant two-body potential whose
Fourier transform is negative definite. \ The main result is that if the
ground state energy $E$ is plotted as a function of the number of particles
$A$, then the function $E(A)$ is convex for even $A$ modulo $2N$.
\ Furthermore $E(A)$ for odd $A$ is bounded below by the average of the two
neighboring even values, $E(A-1)$ and $E(A+1)$. \ This is illustrated in Fig.
\ref{convex} for both the weak attractive and strong attractive cases. \ This
convexity pattern could be regarded as an $SU(2N)$ generalization of even-odd
staggering for the ground state energy in the attractive two-component system.
\

A weaker form of this inequality was proven for $A\leq2N$ and zero-range
attractive interactions \cite{Chen:2004rq}. \ Here we extend the proof to any
$A$ and any $SU(2N)$-invariant potential with a negative-definite Fourier
transform. \ Another difference between this and the previous analysis is that
we use a fixed particle number formalism rather than the grand canonical
formalism. \ This is essential for strongly-attractive interactions and the
requirement of keeping the physics in the low-energy regime. \ By limiting the
number of particles we avoid a collapse towards high densities and the need
for hard core repulsion for stability.

We start by considering $2N$ degenerate components of nonrelativistic fermions
in $d$ spatial dimensions. \ We assume the interactions are governed by an
$SU(2N)$-invariant two-body potential $V(\vec{r})$ whose Fourier transform
$\tilde{V}(\vec{p})$ is strictly negative. \ We also allow an $SU(2N)$%
-invariant external potential $U(\vec{r})$ whose properties are not
restricted. \ The general form of the Hamiltonian is%
\begin{align}
H  &  =-\frac{1}{2m}\sum_{i=1,\cdots,2N}\int d^{d}\vec{r}\;a_{i}^{\dagger
}(\vec{r})\vec{\nabla}^{2}a_{i}(\vec{r})+\int d^{d}\vec{r}\;U(\vec{r}%
)\rho(\vec{r})\nonumber\\
&  +\frac{1}{2}\int d^{d}\vec{r}d^{d}\vec{r}^{\prime}\;\colon\rho(\vec
{r})V(\vec{r}-\vec{r}^{\prime})\rho(\vec{r}^{\prime})\colon,
\end{align}
where $\rho(\vec{r})$ is the $SU(2N)$-invariant density,%
\begin{equation}
\rho(\vec{r})=\sum_{i=1,\cdots,2N}a_{i}^{\dagger}(\vec{r})a_{i}(\vec{r}).
\end{equation}
The $\colon$ symbols denote normal ordering. \ We consider the system on a
hypercubic lattice using a transfer matrix formalism. \ We let $\vec{n}%
=(\vec{n}_{s},n_{t})$ represent $(d+1)$-dimensional lattice vectors. \ The
subscript $s$ on $\vec{n}_{s}$ denotes a $d$-dimensional spatial lattice
vector. \ We write the $d$-dimensional spatial lattice unit vectors as
$\hat{1},\cdots,\hat{d}$. \ Throughout our discussion of the lattice system we
use dimensionless parameters and operators which correspond with physical
values multiplied by the appropriate power of the spatial lattice spacing $a$.
\ We let $a_{t}$ be the temporal lattice spacing and $\alpha_{t}$ be the ratio
$a_{t}/a$. \ $L$ denotes the spatial length of the periodic hypercubic lattice.

We use the notation $\tilde{V}(2\pi\vec{k}_{s}/L)$ for the Fourier transform
of the lattice potential $V(\vec{n}_{s})$,%
\begin{equation}
\tilde{V}(2\pi\vec{k}_{s}/L)=\sum_{\vec{n}_{s}}V(\vec{n}_{s})e^{i2\pi\vec
{n}_{s}\cdot\vec{k}_{s}/L}.
\end{equation}
By assumption $\tilde{V}(2\pi\vec{k}_{s}/L)$ is strictly negative. \ Let $M$
be the normal-ordered transfer matrix operator%
\begin{equation}
M=\colon\exp\left[  -\alpha_{t}H_{\text{free}}-\alpha_{t}\sum_{\vec{n}_{s}%
}U(\vec{n}_{s})\rho(\vec{n}_{s})-\frac{\alpha_{t}}{2}\sum_{\vec{n}_{s},\vec
{n}_{s}^{\prime}}\rho(\vec{n}_{s})V(\vec{n}_{s}-\vec{n}_{s}^{\prime})\rho
(\vec{n}_{s}^{\prime})\right]  \colon,
\end{equation}
where $H_{\text{free}}$ is the free lattice Hamiltonian,%
\begin{equation}
H_{\text{free}}=-\frac{1}{2m}\sum_{\vec{n}_{s}}\sum_{\hat{l}_{s}=\hat
{1},\cdots,\hat{d}}\sum_{i=1,\cdots,2N}\left\{  a_{i}^{\dagger}(\vec{n}%
_{s})\left[  a_{i}(\vec{n}_{s}+\hat{l}_{s})+a_{i}(\vec{n}_{s}-\hat{l}%
_{s})-2a_{i}(\vec{n}_{s})\right]  \right\}  .
\end{equation}
\ Let $V^{-1}(\vec{n}_{s})$ be the inverse of $V(\vec{n}_{s}),$%
\begin{equation}
V^{-1}(\vec{n}_{s})=\frac{1}{L^{d}}\sum_{\vec{k}_{s}}\frac{e^{-i2\pi\vec
{n}_{s}\cdot\vec{k}_{s}/L}}{\tilde{V}(2\pi\vec{k}_{s}/L)}.
\end{equation}
We now rewrite powers of $M$ using an auxiliary field $\phi,$%
\begin{equation}
M^{L_{t}}=%
{\displaystyle\int}
D\phi\;e^{-S(\phi)}\;M_{L_{t}-1}(\phi)\times\cdots\times M_{0}(\phi),
\end{equation}
where%
\begin{equation}
S(\phi)=-\frac{\alpha_{t}}{2}\sum_{n_{t}}\sum_{\vec{n}_{s},\vec{n}_{s}%
^{\prime}}\phi(\vec{n}_{s},n_{t})V^{-1}(\vec{n}_{s}-\vec{n}_{s}^{\prime}%
)\phi(\vec{n}_{s}^{\prime},n_{t}),
\end{equation}%
\begin{equation}
M_{n_{t}}(\phi)\equiv\colon\exp\left[  -\alpha_{t}H_{\text{free}}-\alpha
_{t}\sum_{\vec{n}_{s}}U(\vec{n}_{s})\rho(\vec{n}_{s})+\alpha_{t}\sum_{\vec
{n}_{s}}\phi(\vec{n}_{s},n_{t})\rho(\vec{n}_{s})\right]  \colon,
\end{equation}%
\begin{equation}
D\phi=\prod_{\vec{k}_{s}}\left[  -\tilde{V}(2\pi\vec{k}_{s}/L)\right]
^{-L_{t}/2}\times\prod_{\vec{n}_{s},n_{t}}\frac{d\phi(\vec{n}_{s},n_{t}%
)}{\sqrt{2\pi/\alpha_{t}}}.
\end{equation}

Let $f^{(1)}(\vec{n}_{s}),f^{(2)}(\vec{n}_{s}),\cdots$ be a complete set of
orthonormal real-valued functions of the spatial lattice sites $\vec{n}_{s}$.
\ We refer to these functions as orbitals. \ We take $f^{(1)}(\vec{n}_{s})$ to
be strictly positive but otherwise regard the form for the orbitals to be
arbitrary. \ If the total number of lattice sites is $L^{d}$ then we have a
total of $L^{d}$ orbitals. \ We denote a one-particle state with component $i$
in the $k^{th}$ orbital as $\left\vert f_{i}^{(k)}\right\rangle $.

Let $\mathcal{B}$ and $\mathcal{C}$ be any finite subsets of the orbital
indices, $\mathcal{B},\mathcal{C}\subset\{1,2,\cdots,L^{d}\}$. \ From these we
define $\left\vert \mathcal{B}^{j}\mathcal{C}^{2N-j}\right\rangle $ as the
quantum state where each of $j$ components fill orbitals $\mathcal{B}$ and
each of the remaining $2N-j$ components fill the orbitals $\mathcal{C}$. \ The
order of the component labels is irrelevant, and so we assume that the first
$j$ components fill orbitals $\mathcal{B}$ and last $2N-j$ components fill
orbitals $\mathcal{C}$. \ The total number of fermions in state $\left\vert
\mathcal{B}^{j}\mathcal{C}^{2N-j}\right\rangle $ is $j\left\vert
\mathcal{B}\right\vert +\left(  2N-j\right)  \left\vert \mathcal{C}\right\vert
,$\ where $\left\vert \mathcal{B}\right\vert $ and $\left\vert \mathcal{C}%
\right\vert $ are the number of elements in $\mathcal{B}$ and $\mathcal{C}$ respectively.

We define $E_{\mathcal{B}^{j}\mathcal{C}^{2N-j}}$ as the energy of the lowest
energy eigenstate with nonzero inner product with $\left\vert \mathcal{B}%
^{j}\mathcal{C}^{2N-j}\right\rangle $. \ We let $Z_{\mathcal{B}^{j}%
\mathcal{C}^{2N-j}}^{L_{t}}$ be the expectation value of $M^{L_{t}}$ for
$\left\vert \mathcal{B}^{j}\mathcal{C}^{2N-j}\right\rangle $,%
\begin{equation}
Z_{\mathcal{B}^{j}\mathcal{C}^{2N-j}}^{L_{t}}=\left\langle \mathcal{B}%
^{j}\mathcal{C}^{2N-j}\right\vert M^{L_{t}}\left\vert \mathcal{B}%
^{j}\mathcal{C}^{2N-j}\right\rangle .
\end{equation}
In the limit of large $L_{t}$ the contribution from the lowest energy
eigenstate dominates and therefore%
\begin{equation}
E_{\mathcal{B}^{j}\mathcal{C}^{2N-j}}=-\lim_{L_{t}\rightarrow\infty}\frac
{\ln\left(  Z_{\mathcal{B}^{j}\mathcal{C}^{2N-j}}^{L_{t}}\right)  }{\alpha
_{t}L_{t}}.
\end{equation}
We can write $Z_{\mathcal{B}^{j}\mathcal{C}^{2N-j}}^{L_{t}}$ using the
auxiliary field $\phi$,%
\begin{equation}
Z_{\mathcal{B}^{j}\mathcal{C}^{2N-j}}^{L_{t}}=%
{\displaystyle\int}
D\phi\;e^{-S(\phi)}\;\left\langle \mathcal{B}^{j}\mathcal{C}^{2N-j}\right\vert
M_{L_{t}-1}(\phi)\times\cdots\times M_{0}(\phi)\left\vert \mathcal{B}%
^{j}\mathcal{C}^{2N-j}\right\rangle .
\end{equation}
At this point we define matrix elements for the one particle states,%
\begin{equation}
\mathcal{M}_{k^{\prime},k}(\phi)=\left\langle f_{i}^{(k^{\prime})}\right\vert
M_{L_{t}-1}(\phi)\times\cdots\times M_{0}(\phi)\left\vert f_{i}^{(k)}%
\right\rangle . \label{matrix_K}%
\end{equation}
The component index $i$ in (\ref{matrix_K}) does not matter due to the
$SU(2N)$ symmetry. \ Each entry of the matrix $\mathcal{M}_{k^{\prime},k}%
(\phi)$ is real. \ We let $\mathcal{M}_{\mathcal{B}}(\phi)$ be the $\left\vert
\mathcal{B}\right\vert \times\left\vert \mathcal{B}\right\vert $ submatrix
consisting of the rows and columns in $\mathcal{B}$ and let $\mathcal{M}%
_{\mathcal{C}}(\phi)$ be the $\left\vert \mathcal{C}\right\vert \times
\left\vert \mathcal{C}\right\vert $ submatrix for $\mathcal{C}$.

Each normal-ordered transfer matrix operator $M_{n_{t}}(\phi)$ has only
single-particle interactions with the auxiliary field and no direct
interactions between particles. \ Therefore it follows that
\begin{equation}
Z_{\mathcal{B}^{j}\mathcal{C}^{2N-j}}^{L_{t}}=%
{\displaystyle\int}
D\phi\;e^{-S(\phi)}\;\left[  \det\mathcal{M}_{\mathcal{B}}(\phi)\right]
^{j}\left[  \det\mathcal{M}_{\mathcal{C}}(\phi)\right]  ^{2N-j}.
\label{determinant}%
\end{equation}
This result is perhaps more transparent if we pretend for the moment that each
of the $j\left\vert \mathcal{B}\right\vert +\left(  2N-j\right)  \left\vert
\mathcal{C}\right\vert $ particles carries an extra quantum number which makes
them distinguishable. \ We label the extra quantum number as $X.$ \ So long as
the initial and final state wavefunctions are completely antisymmetric in $X$
for particles of the same component then this error in quantum statistics has
no effect on the final amplitude. \ So we can factorize the transfer matrices
$M_{n_{t}}(\phi)$ as a product of transfer matrices for each $X$. \ This then
leads directly to (\ref{determinant})$.$

Let $n_{1}$ and $n_{2}$ be integers such that $0\leq2n_{1}<j<2n_{2}\leq2N.$
\ Let us define the new positive-definite measure%
\begin{equation}
\tilde{D}\phi=D\phi\;e^{-S(\phi)}\;\left[  \det\mathcal{M}_{\mathcal{B}}%
(\phi)\right]  ^{2n_{\text{$1$}}}\left[  \det\mathcal{M}_{\mathcal{C}}%
(\phi)\right]  ^{2N-2n_{2}}, \label{redefined_measure}%
\end{equation}
so that%
\begin{equation}
Z_{\mathcal{B}^{j}\mathcal{C}^{2N-j}}^{L_{t}}=%
{\displaystyle\int}
\tilde{D}\phi\;\left[  \det\mathcal{M}_{\mathcal{B}}(\phi)\right]
^{j-2n_{\text{$1$}}}\left[  \det\mathcal{M}_{\mathcal{C}}(\phi)\right]
^{2n_{2}-j}.
\end{equation}
Then by the H\"{o}lder inequality $\left\vert Z_{\mathcal{B}^{j}%
\mathcal{C}^{2N-j}}^{L_{t}}\right\vert $ is bounded above by%
\begin{align}
&  \left[
{\displaystyle\int}
\tilde{D}\phi\;\left\vert \det\mathcal{M}_{\mathcal{B}}(\phi)\right\vert
^{2n_{\text{$2$}}-2n_{\text{$1$}}}\right]  ^{\frac{j-2n_{\text{$1$}}%
}{2n_{\text{$2$}}-2n_{\text{$1$}}}}\left[
{\displaystyle\int}
\tilde{D}\phi\;\left\vert \det\mathcal{M}_{\mathcal{C}}(\phi)\right\vert
^{2n_{\text{$2$}}-2n_{\text{$1$}}}\right]  ^{\frac{2n_{\text{$2$}}%
-j}{2n_{\text{$2$}}-2n_{\text{$1$}}}}\nonumber\\
&  =\left(  Z_{\mathcal{B}^{2n_{2}}\mathcal{C}^{2N-2n_{2}}}^{L_{t}}\right)
^{\frac{j-2n_{\text{$1$}}}{2n_{\text{$2$}}-2n_{\text{$1$}}}}\left(
Z_{\mathcal{B}^{2n_{1}}\mathcal{C}^{2N-2n_{1}}}^{L_{t}}\right)  ^{\frac
{2n_{\text{$2$}}-j}{2n_{\text{$2$}}-2n_{\text{$1$}}}}.
\end{align}
Taking the limit $L_{t}\rightarrow\infty$ we deduce that the energies satisfy
the inequality%
\begin{equation}
E_{\mathcal{B}^{j}\mathcal{C}^{2N-j}}\geq\tfrac{j-2n_{\text{$1$}}%
}{2n_{\text{$2$}}-2n_{\text{$1$}}}E_{\mathcal{B}^{2n_{2}}\mathcal{C}%
^{2N-2n_{2}}}+\tfrac{2n_{\text{$2$}}-j}{2n_{\text{$2$}}-2n_{\text{$1$}}%
}E_{\mathcal{B}^{2n_{1}}\mathcal{C}^{2N-2n_{1}}}. \label{inequality}%
\end{equation}
This is a statement of convexity for $E_{\mathcal{B}^{j}\mathcal{C}^{2N-j}}$
as a function of $j$ between even endpoints $j=2n_{\text{1}}$ and
$j=2n_{\text{2}}$. \ If we now take $\left\vert \mathcal{B}\right\vert =K+1$
and $\left\vert \mathcal{C}\right\vert =K$, then the total particle number is
$A=2NK+j$ and $A$ lies between $2NK$ and $2N(K+1)$. \ The inequality in
(\ref{inequality}) is precisely the convexity pattern in Fig. \ref{convex} for
$E(A)$ as a function of particle number.

We point out that for the special case $K=0$, we can take $\mathcal{B}$ to be
the first orbital and $\mathcal{C}$ to be the empty set. \ In this case
$\mathcal{M}_{\mathcal{B}}(\phi)$ is simply a number. \ Furthermore since
$f^{(1)}(\vec{n}_{s})$ is strictly positive, $\mathcal{M}_{\mathcal{B}}(\phi)$
is also positive so long as the temporal lattice step $a_{t}$ is not
excessively large. \ Since $\det\mathcal{M}_{\mathcal{B}}(\phi)=\mathcal{M}%
_{\mathcal{B}}(\phi)>0$ it is no longer necessary that the power of
$\det\mathcal{M}_{\mathcal{B}}(\phi)$ be even to insure positivity.
\ Therefore $E(A)$ is actually convex for all $A$ between $0$ and $2N$ and not
just even $A$.

\begin{figure}
[ptb]
\begin{center}
\includegraphics[
height=3.0882in,
width=3.224in
]%
{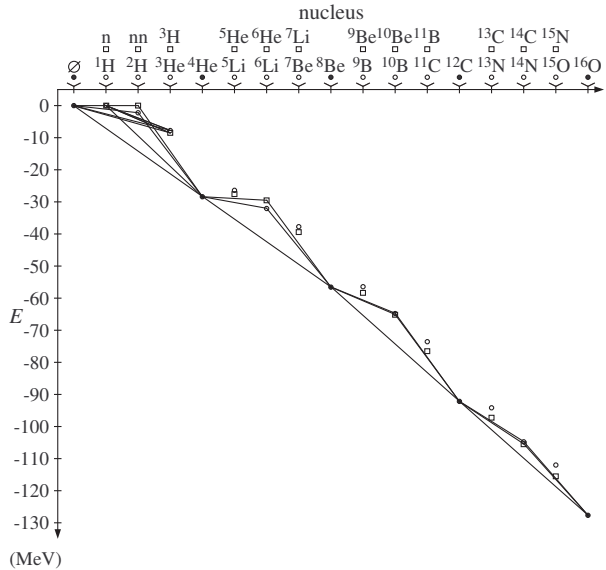}%
\caption{Plot of the energy versus particle number for the most stable light
nuclei with up to $16$ nucleons. \ The line segments show the convexity lower
bounds.}%
\label{su4_all}%
\end{center}
\end{figure}

These convexity relations could be checked using any number of attractive
$SU(2N)$ models in various dimensions. \ This will be checked in future
studies. \ Here we instead examine actual nuclear physics data to investigate
Wigner's approximate $SU(4)$ symmetry in light nuclei. \ It is by no means
clear that the interactions of nucleons in light nuclei can be approximately
described by an attractive $SU(4)$-symmetric potential. \ Recent results from
nuclear lattice simulations hint that this might be possible
\cite{Borasoy:2005yc,Borasoy:2006qn}, however there are forces even at lowest
order in chiral effective field theory which break $SU(4)$ invariance in
addition to being repulsive. \ Nevertheless all of the $SU(4)$ convexity
constraints are in fact satisfied for the most stable light nuclei with up to
$16$ nucleons as can be seen in Fig. \ref{su4_all}. \ The line segments drawn
show all of the convexity lower bounds.%

There have been several recent studies of alpha clustering in nuclear matter
\cite{Ropke:1998qs,Tohsaki:2001an,Funaki:2002fn,Schuck:2003ep,Yamada:2003cz,Sedrakian:2004fh}
as well as multiparticle clustering in other systems
\cite{Schlottmann:1994JPCM,Kamei:2005JPS,Lecheminant:2005PRL,Wu:2005PRL,Capponi:2006,Wu:2006MPLB}%
. \ The results presented here give sufficient conditions for the onset of
this multiparticle clustering phase. \ One can also make a definite prediction
about the $j$-component quasiparticle energy gaps. \ Starting from a
$2NK$-fermion $SU(2N)$-symmetric state, let $\delta_{j}$ be the extra energy
required per fermion to add $j$ fermions, all of different components. \ The
ground state energy for $2NK+j$ fermions is a convex function for even $j$ in
the interval from $j=0$ to $j=2N$. \ Therefore it follows that $\delta_{2}%
\geq\delta_{4}\geq\cdots\geq\delta_{2N}$. \ Since the ground state energy for
$2NK+j$ fermions is also convex for $j=0,1,2$, we conclude furthermore that
$\delta_{1}\geq\delta_{2}\geq\delta_{4}\cdots\geq\delta_{2N}$. \ We note that
for the strongly attractive case these energy gaps are negative, and it is
more natural to speak of energy gaps per missing fermion for the corresponding
$j$-component quasiholes$,\delta_{j}^{h}$. \ In this case we find again
$\delta_{1}^{h}\geq\delta_{2}^{h}\geq\delta_{4}^{h}\cdots\geq\delta_{2N}^{h}$.

In summary we have derived a general result on spectral convexity with respect
to particle number for $2N$ degenerate components of fermions. \ We assume
only that the interactions are governed by an $SU(2N)$-invariant two-body
potential whose Fourier transform is negative definite. \ The ground state
energy $E$ as a function of the number of particles $A$ is convex for even $A$
modulo $2N$. \ Also $E(A)$ for odd $A$ is bounded below by the average of the
two neighboring even values, $E(A-1)$ and $E(A+1)$. \ When applied to light
nuclei for $A\leq16$ all of the convexity bounds for $SU(4)$ are satisfied.
\ These results give further evidence that an approximate description of light
nuclei may be possible using an attractive $SU(4)$-symmetric potential. \ This
would be a direction worth pursuing since the same theory could then be
applied to dilute neutron-rich matter with a finite number of protons. \ The
residual $SU(2)\times SU(2)$ symmetry for proton spins and neutron spins would
guarantee that the Monte Carlo simulation could be done without fermion sign
oscillations. \ The physics of this quantum system would be helpful in
understanding the superfluid properties of dilute neutron-rich matter in the
inner crust of neutron stars.

\textit{The author thanks Gautam Rupak and Thomas Sch\"{a}fer for discussions
and correspondence with Philippe Lecheminant and Congjun Wu. \ This work was
supported in part by DOE grant DE-FG02-03ER41260.}

\bibliographystyle{apsrev}
\bibliography{NuclearMatter}

\end{document}